\documentclass[12pt]{article} 
\usepackage{epsf,multicol,ifthen}
\usepackage{graphicx,amstext,amssymb}

\begin{document}
\begin{center}
{\Large \bf Integral and derivative dispersion relations
in the analysis of the data on $pp$ and $\bar pp$ forward scattering.}\\
\vskip 1.cm
 J.R. Cudell$^{\dagger}$\footnote{E-mail: JR.Cudell@ulg.ac.be},
 E. Martynov$^{\dagger,\#}$\footnote{E-mail: E.Martynov@guest.ulg.ac.be, martynov@bitp.kiev.ua},
 O. Selyugin$^{\dagger,*}$\footnote{E-mail: Selugin@nuclth20.phys.ulg.ac.be, selugin@tsun1.jinr.ru}
\end{center}

\noindent {\small $^{\dagger}$Institut de Physique, Universit\`{e} de
Li\`{e}ge, B\^{a}t.
B5-a, Sart Tilman, B4000 Li\`{e}ge, Belgium\\
$^{\#}$Bogolyubov Institute for Theoretical Physics,
Metrogocheskaya str. 14b,  Kiev 03143\\
$^{*}$Bogoliubov Theoretical Laboratory, Joint Institute of Nuclear
Research, 141980, Dubna,\\  Moscow region, Russia.}

\begin{center}
{\bf Abstract}
\end{center}

Integral and derivative dispersion relations (DR) are considered for the
$pp$ and $\bar pp$ forward scattering amplitudes. A new representation for
the derivative DR, valid at lower energies than the standard one, is
obtained. The data on the total cross sections of $pp (\bar pp)$
interaction as well as those on the ratio of the real part to the
imaginary part of the forward amplitude are analyzed within various forms
of the DR and high-energy Regge models. It is shown that three models for
the pomeron, simple pole with intercept larger than one, triple pole
pomeron and double pole pomeron (both with intercept equal to one) lead to
practically equivalent descriptions of the data at $\sqrt{s}>5$ GeV. It is
also shown that the low-energy part of the dispersion integral (from the
two-proton threshold up to $\sqrt{s}=5$ GeV) allows one to reproduce well
the data on $\rho$ at lower energies without additional free parameters.

\section{Introduction}
The energy dependence of the hadronic total cross sections as well as that
of the parameters $\rho=\Re eA(s,0)/\Im mA(s,0)$ - the ratios of the real to
the imaginary part of the forward scattering amplitudes - was widely discussed
quite a long time ago (see \cite {rhohist,DDR1,DDR2,KN} and references
therein). However, in spite of recent detailed investigations
on the subject, the theoretical situation remains somewhat undecided,
mainly because of the $\rho$ parameter.

In the papers \cite{COMPETE}, all available data on $\sigma_{tot}(s)$ and
$\rho(s)$ for hadron-hadron, photon-hadron and photon-photon interactions
were considered. Many analytical models for the forward scattering
amplitudes were fitted and compared. The ratio $\rho$ was calculated in
explicit form, from the imaginary part parametrised explicitly
by contributions from the pomeron and secondary reggeons. The values of the free
parameters were determined from the fit to the data at $s\ge s_{min}$,
where $\sqrt{s_{min}}=5$ GeV. Omitting all details, we note here the main two
conclusions. The best description of the data (with the minimal
$\chi^{2}/dof$, where $dof$ is the number of degrees of freedom) is obtained
for the model with $\sigma_{tot}$ rising as $\log^{2}s$. The model with
$\sigma_{tot}(s)\propto s^{\epsilon}, \epsilon >0$ was excluded from the
list of the best models (in accordance with COMPETE criteria, see details
in \cite{COMPETE}).

Analysis of these results shows that they are due to a poor
description of $\rho$ data at low energy. On the
other hand, there are a few questions concerning the explicit Regge-type
models usually used for analysis and description of the data. How low in
energy can the Regge parametrisations be extended, as they are written as
functions of the asymptotic variable $s$ rather than the "Regge" variable
$\cos\theta_{t}$ ($=E/m$ in the laboratory system for identical colliding
particles)? At which energies can the "asymptotic" normalization
\begin{equation}\label{eq:asympt. norm.}
\sigma_{tot}(s)=\frac{1}{s}\Im mA(s,0)
\end{equation}
instead of the standard one
\begin{equation}\label{eq:stand. norm.}
\sigma_{tot}(s)=\frac{1}{2mE}\Im mA(s,0)
\end{equation}
be used? And last, how much do the analytic expressions for $\rho$
based on the derivative dispersion relations deviate from those calculated
in the integral form?

In this paper, we try to answer these questions considering three pomeron
models (simple pole in the complex-$j$ plane with intercept above unity,
double pole and triple
pole with intercepts equal to unity) for $pp$ and $\bar pp$ interactions
at $\sqrt{s}\geq 5$ GeV.

\section{Integral and derivative dispersion relations.}
The amplitude $A(s,t,u)$ is an analytic function of its variables.
Consequently it must satisfy the integral dispersion relation (IDR)
which can be derived
from the Cauchy theorem on analytic functions. Generally, for
proton-proton and proton-antiproton amplitudes, two subtractions must be
made. However, assuming, in accordance with many analyses, that the odderon
does not contribute asymptotically, one can show that the dispersion
relations for $pp$ and $\bar pp$ amplitudes can be reduced to those with
one subtraction constant \cite{rhohist}:
\begin{equation}\label{eq:dispers II}
\rho_{\pm}\sigma_{\pm}=\frac{B}{2m_{p}p}\,\, +\frac{E}{\pi p}\,{\rm
P}\!\int\limits_{m_{p}}^{\infty}\left
[\frac{\sigma_{\pm}}{E'(E'-E)}-\frac{\sigma_{\mp}}{E'(E'+E)}\right ]p'\,
dE'
\end{equation}
where $m_{p}$ is the proton mass, $E$ and $p$ are the energy and momentum
of the proton in the laboratory system, and $B$ is a subtraction constant,
usually determined from the fit to the data. The indices
$+(-)$ stand respectively for the
$pp$ and $ (\bar pp)$ amplitudes. The standard normalization (\ref{eq:stand.
norm.}) is chosen in Eq.(\ref{eq:dispers II}).

In the above expression, the pole contributions and the part of the
integral over the unphysical cut from the two-pion to the two-proton
threshold are omitted because they are $\lesssim 1\%$ (see, e.g.
\cite{valeng}) in the region of interest ($\sqrt{s}\ge 5$ GeV). In
other words, we shall investigate here only the contribution of the physical
region to the dispersion integral.

The derivative dispersion relations (DDR) were obtained \cite{DDR1,DDR2}
separately for crossing-even and crossing-odd amplitudes
\begin{equation}\label{eq:crossampl}
f_{\pm}(s,0)=A_{+}(s,0)\pm A_{-}(s,0).
\end{equation}
They are very useful in a practice due to their simple analytical form at
high energies, $E\gg m_{p}$:
\begin{equation}\label{eq:ddrev-as}
\Re ef_{+}(E,0)\approx (E/m_{p})^{\alpha}\tan \left [ \frac{\pi}{2}\left
(\alpha-1+E\frac{d}{dE}\right )\right ] \Im
mf_{+}(E,0)/(E/m_{p})^{\alpha}.
\end{equation}
However it is important to estimate the corrections to these
asymptotic relations (\ref{eq:ddrev-as}) if one is to use them at finite
$s$.

The starting point is the dispersion integral relation for an even
amplitude $f_{+}$ (with poles and subtraction constant omitted $-$ they will
be re-introduced at the end of calculations $-$). The amplitude is normalized
so that $2m_{p}p_{\, lab}\sigma_{t}^{+}=\Im mf_{+}(E,0)$.
Let us consider the relation at $t=0$:
\begin{equation}\label{eq:dispint0}
\Re ef_{+}(E,0)=\frac{2E^{2}}{\pi}{\rm P}\int\limits_{m_{p}}^{\infty}\,
\frac{dE'}{E'(E'^{2}-E^{2})}\Im mf_{+}(E',0)
\end{equation}
and represent it in the form
$$
\frac{2E^{2}}{\pi}{\rm P}\int\limits_{m_{p}}^{\infty}\, \frac{dE'\,
(E'/m_{p})^{\alpha}}{E'(E'^{2}-E^{2})}g(E'),
$$
where $g(E')=\Im mf_{+}(E',0)/(E'/m_{p})^{\alpha}$ and $\alpha $ is an
arbitrary constant (it is not an intercept!).

Integrating by parts (and using the notations
$F(E')=(E'/m_{p})^{\alpha-1}g(E')$, $F'(E)\equiv {dF\over dE}$ for all
functions), we obtain, assuming that $\Im mf_{+}(m_{p},0)=0$:
$$
\Re ef_{+}(E,0)=\frac{E}{\pi m_{p}}  \int\limits_{m_{p}}^{\infty}
dE'\ln\left (\frac{E'+E}{|E'-E|}\right )F'(E').
$$
Now let us transform the integrand functions to the forms
$$
F'(E')=\frac{d}{dE'}\left [ (E'/m_{p})^{\alpha-1}\frac{\Im
mf_{+}(E',0)}{(E'/m_{p})^{\alpha}}\right
]=\frac{1}{m_{p}}(E'/m_{p})^{\alpha-2}\left
(\alpha-1+\frac{d}{d\xi'}\right )g(\xi'),
$$
$$
\ln\frac{E'+E}{|E'-E|}=\ln\frac{e^{\xi'}+e^{\xi}}{|e^{\xi'}-e^{\xi}|}=
\ln\frac{e^{(\xi'-\xi)/2}+e^{-(\xi'-\xi)/2}}{|e^{(\xi'-\xi)/2}-e^{-(\xi'-\xi)/2}|}
=\ln|\coth\frac{1}{2}(\xi'-\xi)|,
$$
where $\xi '=\ln (E'/m_{p}) $ and $\xi =\ln (E/m_{p}) $.

Taking into account the above expression, one can write the integral
(\ref{eq:dispint0}) in the form
\begin{equation}\label{eq:dispint1}
\begin{array}{lll}
\Re ef_{+}(E,0)&=&\frac{E}{\pi m_{p}}\bigg \{ \ln\left
(\frac{E+m_{p}}{E-m_{p}}\right )\Im
mf_{+}(m_{p},0)+\\&+&\int\limits_{0}^{\infty}d\xi'e^{(\alpha-1)\xi'}
\ln\left (|\coth\frac{1}{2}(\xi'-\xi)|\right ) \left
(\alpha-1+\frac{d}{d\xi'}\right )g(\xi')\bigg \}.
\end{array}
\end{equation}

Rewriting the integrand logarithm (at $x=\xi'-\xi \neq 0$) in
(\ref{eq:dispint1})
$$
\ln\bigg | \coth \frac{1}{2}x\bigg |=\left \{
\begin{array}{lll}
&\ln \frac{1+e^{-x}}{1-e^{-x}},\quad &x>0,\\
&\ln \frac{1+e^{x}}{1-e^{x}},\quad &x<0
\end{array} \right .=\ln\frac{1+e^{-|x|}}{1-e^{-|x|}}
=2\sum\limits_{p=0}^{\infty}\frac{e^{-(2p+1)|x|}}{2p+1}.
$$
and expanding other factors in the integrand of (\ref{eq:dispint1}) in
powers of $\xi'-\xi$  (it does not depend on $\xi$, this is just a trick
allowing explicitly to perform the integration and to present the result
as a series)\footnote{Note that it is not allowed if $\xi\to 0$.}
$$
\left (\alpha-1+\frac{d}{d\xi'}\right )g(\xi')\equiv \tilde
g(\xi')=\sum\limits_{k=0}^{\infty}\frac{(\xi'-\xi)^{k}}{k!}D_{\xi}^{k}\tilde
g(\xi),
$$
where $D_{\xi}=\frac{d}{d\xi}$ and $\xi \neq 0,$
$$
e^{(\alpha-1)\xi'}=e^{(\alpha-1)\xi}e^{(\alpha-1)(\xi'-\xi)}=e^{(\alpha-1)\xi}
\sum\limits_{n=0}^{\infty}\frac{(\alpha-1)^{n}}{n!}(\xi'-\xi)^{n},
$$
one can write
$$
\Re ef_{+}(E,0)= \frac{2E}{\pi
m_{p}}e^{(\alpha-1)\xi}\sum\limits_{p=0}^{\infty}\frac{1}{2p+1}
\sum\limits_{k=0}^{\infty}\sum\limits_{n=0}^{\infty}\frac{(\alpha-1)^{n}}{k!n!}\cdot
{\cal I}(\xi;p,k,n)\cdot D_{\xi}^{k}\tilde g(\xi)
$$
where
\begin{equation}\label{eq:Iint}
\begin{array}{lll}
{\cal
I}(\xi;p,k,n)&=&\int\limits_{0}^{\infty}d\xi'e^{-(2p+1)|\xi'-\xi|}(\xi'-\xi)^{k+n}\\&=&
\frac{1}{(2p+1)^{k+n+1}}\left
[\Gamma(k+n+1)+(-1)^{k+n}\gamma(k+n+1,\xi)(2p+1)\right ]
\end{array}
\end{equation}
and $\gamma(a,x)$ is the incomplete gamma function. We should note that
one can prove that the asymptotic expression for DDR as well as the
corrections to them do not depend on the auxiliary parameter $\alpha$ (thus
it is unreasonable to determine the parameter $\alpha$ from a fit to
data). So we may put, for convenience, $\alpha=1$. The first term in
(\ref{eq:Iint}) gives the asymptotic form (\ref{eq:ddrev-as})
\cite{DDR1,DDR2,KN} written at $\alpha=1$.

After some transformations we obtain for the corrections the series
expansion in powers of $m_{p}/E$
\begin{equation}\label{eq:ddrcor}
\Re ef_{+}^{(cor)}(E,0)=-\frac{2E}{\pi
m_{p}}\sum\limits_{p=0}^{\infty}\frac{C_{+}(p)}{2p+1} e^{ -\xi(2p+1)}.
\end{equation}
where
$$
C_{+}(p)=\frac{e^{-\xi D_{\xi}}}{2p+1+D_{\xi}}[ \Im mf_{+}(E,0)-E\Im
mf'_{+}(E,0)].
$$
It should be noted that despite an apparent dependence on
$\xi=\ln(E/m_{p})$, the above expression for $C_{+}(p)$ in fact does not
depend on $E$. This can be proven using the properties of $\exp(-\xi
D_{\xi})$.

Then, collecting all terms and adding a subtraction constant $B_{+}$, we
obtain the final expression for $\Re ef_{+}(E,0)$
\begin{equation}\label{eq:ddreven}
\Re ef_{+}(E,0)= B_{+}+E\tan \left [ \frac{\pi}{2}E\frac{d}{dE}\right
]\frac{\Im mf_{+}(E,0)}{E}\,
 -\frac{2}{\pi
}\sum\limits_{p=0}^{\infty}\frac{C_{+}(p)}{2p+1}\left
(\frac{m_{p}}{E}\right )^{2p}.
\end{equation}
A similar expression is obtained for the crossing-odd part of the
amplitudes. If a contribution of asymptotic odderon is absent
then
\begin{equation}\label{eq:ddrodd-2}
\Re ef_{-}(E,0)=-E\cot \left [ \frac{\pi}{2}\frac{Ed}{dE}\right ]\frac{\Im
mf_{-}(E,0)}{E}\,- \frac{2}{\pi
}\sum\limits_{p=0}^{\infty}\frac{C_{-}(p)}{2p+1}\left
(\frac{m_{p}}{E}\right )^{2p+1},
\end{equation}
where
$$
C_{-}(p)=e^{-\xi D_{\xi}}\left \{ \frac{1}{2p+D_{\xi}}\Im
mf'_{-}(E,0)\right \}.
$$

\section{Phenomenology.}
Our aim is to compare the fits of three pomeron models with $\rho(s)$
calculated by two methods. The first one is the integral dispersion
relation, and the second one is the asymptotic form of the derivative
dispersion relation with a subtraction constant. We would like evaluate
the importance of
the normalization on the quality of the
descriptions and on the values of the model parameters.

There are a few difficulties in an analysis of the available experimental
data when the integral dispersion relations are applied.
\begin{itemize}
 \item The imaginary part of the amplitude under consideration must be
known in the whole kinematic region for a given process, starting from
threshold.
 \item It is evident that the known parameterization of the cross
sections, valid at $s\geq s_{min}$, cannot be used for the part of dispersion
integral from the threshold $s_{th}$ to $s_{min}$.
 \item The available experimental data
on the parameters $\rho$ for all processes, $pp, \bar pp, \pi^{\pm}p$ and
$K^{\pm} p$ are of poor quality compared to those for total cross sections.
\end{itemize}

In this paper, we apply as a first step the IDR and DDR only to
$pp$ and $\bar pp$ data. We have fitted the high-energy models to the data
at $\sqrt{s}\ge 5$ GeV.

\subsection{Low-energy data.}\label{sec:low}
Low-energy total cross sections for $pp$ (143 points) and $\bar pp$ (220
points) interactions at $\sqrt{s}<5$ GeV were parameterized as functions
of the proton momentum in the laboratory system by different expressions
for different intervals of momentum.

The values of the free parameters are determined from a fit to the data
\cite{data} at $s<s_{min}$ with the only constraint that the cross sections
calculated from a fit to the data at $s<s_{min}$ must coincide at
$s=s_{min}$ with those given by the fit to a specific high-energy
model, valid for $s>s_{min}$ (all data on $\sigma$ and $\rho$
are taken from \cite{data}).

Thus we perform an overall fit in three steps.
\begin{enumerate}
\item
 The chosen model for high-energy cross-sections is fitted to the data on
the cross sections only (without $\rho$ data) at $s>s_{min}$.
 \item The obtained "high-energy" parameters are fixed. The "low-energy"
pa\-ra\-meters are determined from the fit at $s<s_{min}$, but with
$\sigma_{pp}^{\bar pp}(s_{min})$  given by the first step.
 \item
The subtraction constant $B_{+}$ is determined from the fit at $s>s_{min}$
with all other parameters kept fixed.
\end{enumerate}

Then, without fitting, we calculate the ratios $\rho_{pp}$ and $\rho_{\bar
pp}$ at all energies above the physical threshold. The results of such
procedure (for the cross sections) are given in Fig.1 and Fig.2, and will
be detailed in a forthcoming paper.

\subsection{High energy. Pomeron models.}\label{sec.high}

As an example, we consider three models leading to different asymptotic
behaviors for the total cross sections. For each model, we investigate how
the integral and derivative dispersion relations work. We start from the
explicit parameterization of the total $pp$ and $\bar pp$ cross-sections,
then, to find the ratios of the real to imaginary parts, we apply the IDR
making use of the second method for a calculation of the low-energy part
of the dispersion integral. Then we compare results for ratios calculated
through the DDR.

All the above-mentioned models include the contributions of pomeron, $f$
and $\omega$ reggeons (we consider these reggeons as effective ones
because it is not reasonable to add other secondary reggeons provided only
the $pp$ and $\bar pp$ data are fitted.)

\begin{equation}\label{eq:sigmod}
\sigma_{pp}^{\bar pp}={\cal P}(E)+R_{f}(E)\pm R_{\omega}(E),
\end{equation}
where $R_{f}=R_{+}, R_{\omega}=R_{-}$ and
\begin{equation}\label{eq:reggeon}
R_{\pm}(E)=g_{\pm}\left (\frac{E}{\lambda m_{p}}\right
)^{\alpha_{\pm}(0)-1}.
\end{equation}
The parameter $\lambda$ can play a role only for the pomeron term in the
triple pole model (see below), in the simple pole and dipole models
$\lambda =1$ but we keep it to have a common form and notation for
the three models.

When the imaginary part of amplitude is integrated in IDR we consider (for
comparison) two kinds of normalization: the standard one defined in Eq.
(\ref{eq:stand. norm.}) and the asymptotic one given by Eq.
(\ref{eq:asympt. norm.}). For the latter case, in the expressions for
cross-sections, the replacement $E/\lambda m_{p}\to s/s_{1}$ is made.

Besides, we compare our results with the models which
are written as functions of $-is$ and with the asymptotic normalization
(\ref{eq:asympt. norm.}). We denote such  fits as "$-is$" fits.

\subsubsection{Simple pole pomeron model (SP).}

In this model, the intercept of the pomeron trajectory is larger than unity
\cite{d-l}
\begin{equation}\label{simpleE}
{\cal P}(E)=g\left (\frac{E}{\lambda m_{p}}\right )^{\alpha_{\cal
P}(0)-1}.
\end{equation}
We present the results of the fits using derivative dispersion relations
(with a subtraction constant) and of standard fits with amplitudes defined in
accordance with "$-is$" rule:
\begin{equation}\label{simpleS}
\frac{1}{s}A^{\bar pp}_{pp}(s,0)=ig(-is/s_{1})^{\alpha_{\cal
P}(0)-1}+ig_{+}(-is/s_{1})^{\alpha_{+}(0)-1}\pm
g_{-}(-is/s_{1})^{\alpha_{-}(0)-1}.
\end{equation}

\subsubsection{Dipole pomeron model (DP).}
The pomeron in this model is a double pole in the complex angular momentum
plane with intercept $\alpha_{{\cal P}}(0)=1$.
\begin{equation}\label{doubleE}
{\cal P}(E)=g_{1}+g_{2}\ln(E/\lambda m_{p}).
\end{equation}
For the "$-is$" fit we use
\begin{equation}\label{doubleS}
\frac{1}{s}A^{\bar pp}_{pp}(s,0)=ig_{1}+ig_{2}\ln(-is/s_{1})
+ig_{+}(-is/s_{1})^{\alpha_{+}(0)-1}\pm
g_{-}(-is/s_{1})^{\alpha_{-}(0)-1}.
\end{equation}

\subsubsection{Tripole pomeron model (TP)}
The dominant term at high energy in this model is the contribution of the
hardest $j$-plane Regge singularity allowed by unitarity: the triple pole
at $t=0$ and $j=1$. The form of this contribution is
\begin{equation}\label{tripoleE}
{\cal P}(E)=g_{1}+g_{2}\ln^{2}(E/\lambda m_{p}).
\end{equation}
The prescription "$-is$" gives
\begin{equation}\label{tripoleS}
\frac{1}{s}A^{\bar pp}_{pp}(s,0)=ig_{1}+ig_{2}\ln^{2}(-is/s_{1})
+ig_{+}(-is/s_{1})^{\alpha_{+}(0)-1}\pm
g_{-}(-is/s_{1})^{\alpha_{-}(0)-1}.
\end{equation}

\section{Results, discussion and conclusions.}

Omitting many details of the fits, as well as the numerical values of the
parameters, and postponing them for a forthcoming paper, we concentrate now
on the main results and conclusions. The quality of the
fits is presented in the Table, where $\chi^{2}/N_{p}$ ($N_{p}$ is the
number of experimental points). In Figs. 3 and 4, we show the curves only
for the simple-pole pomeron model. For other models the curves are practically
the same in the region where data are available.

\begin{table}[h]
  \centering
  \caption{The values of $\chi^{2}$ obtained in the
various pomeron models and through the different methods for the
calculation of the ratio~$\rho$.}\label{tab:chi2}

\medskip
\begin{tabular}{|c|c|c|c|c|} \hline
& \multicolumn{2}{c|}{Standard normalization} & \multicolumn{2}{|c|}{High-energy normalization}\\
\cline{2-5}
  & IDR & DDR & IDR & $"-is", B=0$ \\
\hline
 Simple Pole & 1.0462 & 1.0646 & 1.1120 & 1.1209 \\
\hline
 Double Pole & 1.0532 & {1.0454} & 1.1319 & {1.0793} \\
\hline
 Triple Pole & 1.0438 & 1.0456 & 1.1114 & 1.1153 \\
\hline
\end{tabular}
\end{table}

As one can see from the Table, all models give good descriptions of the
data on $\sigma_{tot}$ and $\rho$. Evidently, the fit with the integral
dispersion relations and with the standard normalization is preferable.
While the data on $\sigma$ are described with $\chi^{2}/dof \approx 0.91$,
the data on $\rho$ are described less well, with a $\chi^{2}/N_{p}\approx
1.5$. However we think that this occurs because of the bad quality of the
$\rho$ data.

\begin{figure}[h]
\centering
\includegraphics[scale=0.45]{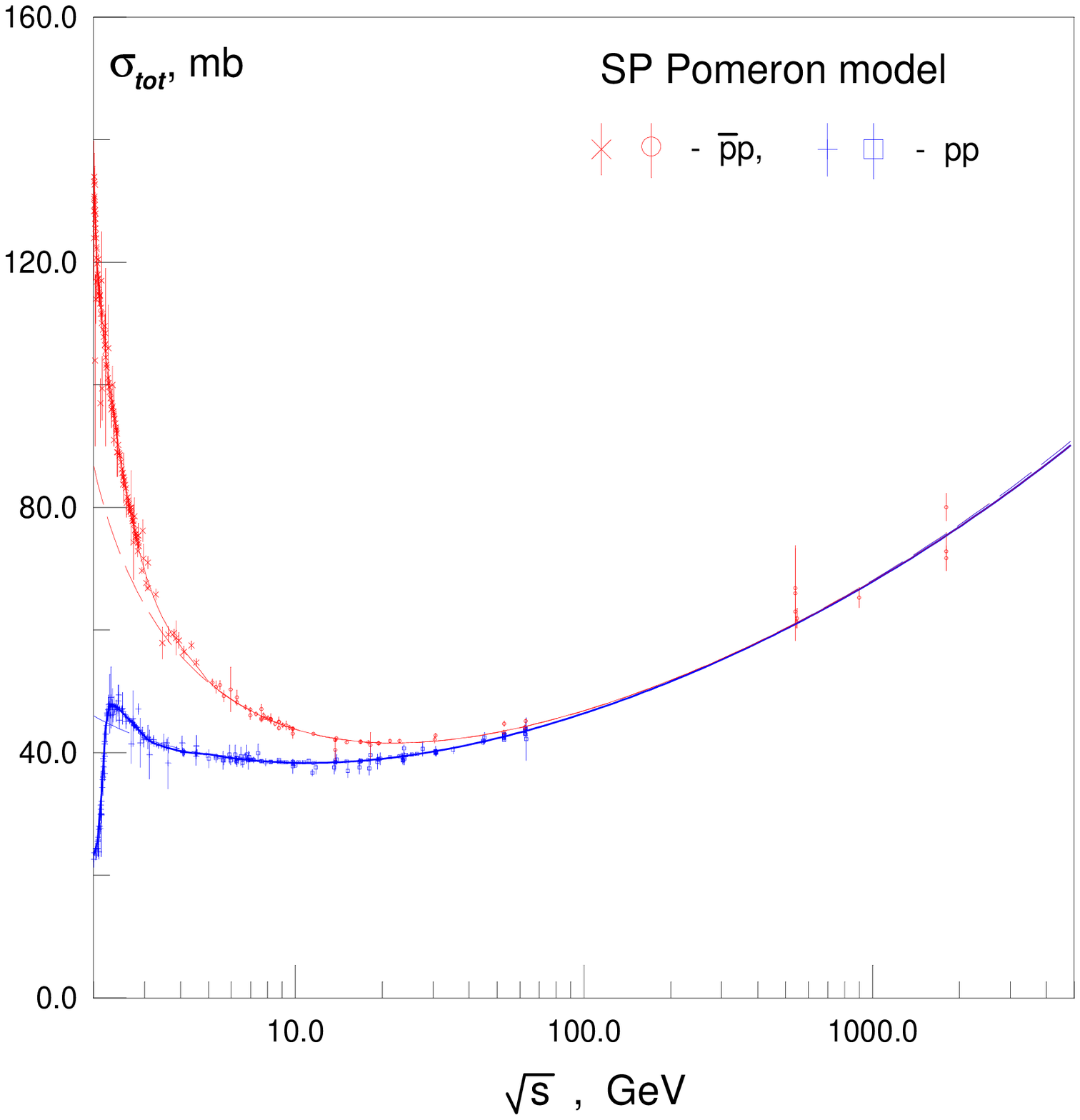}
\includegraphics[scale=0.45]{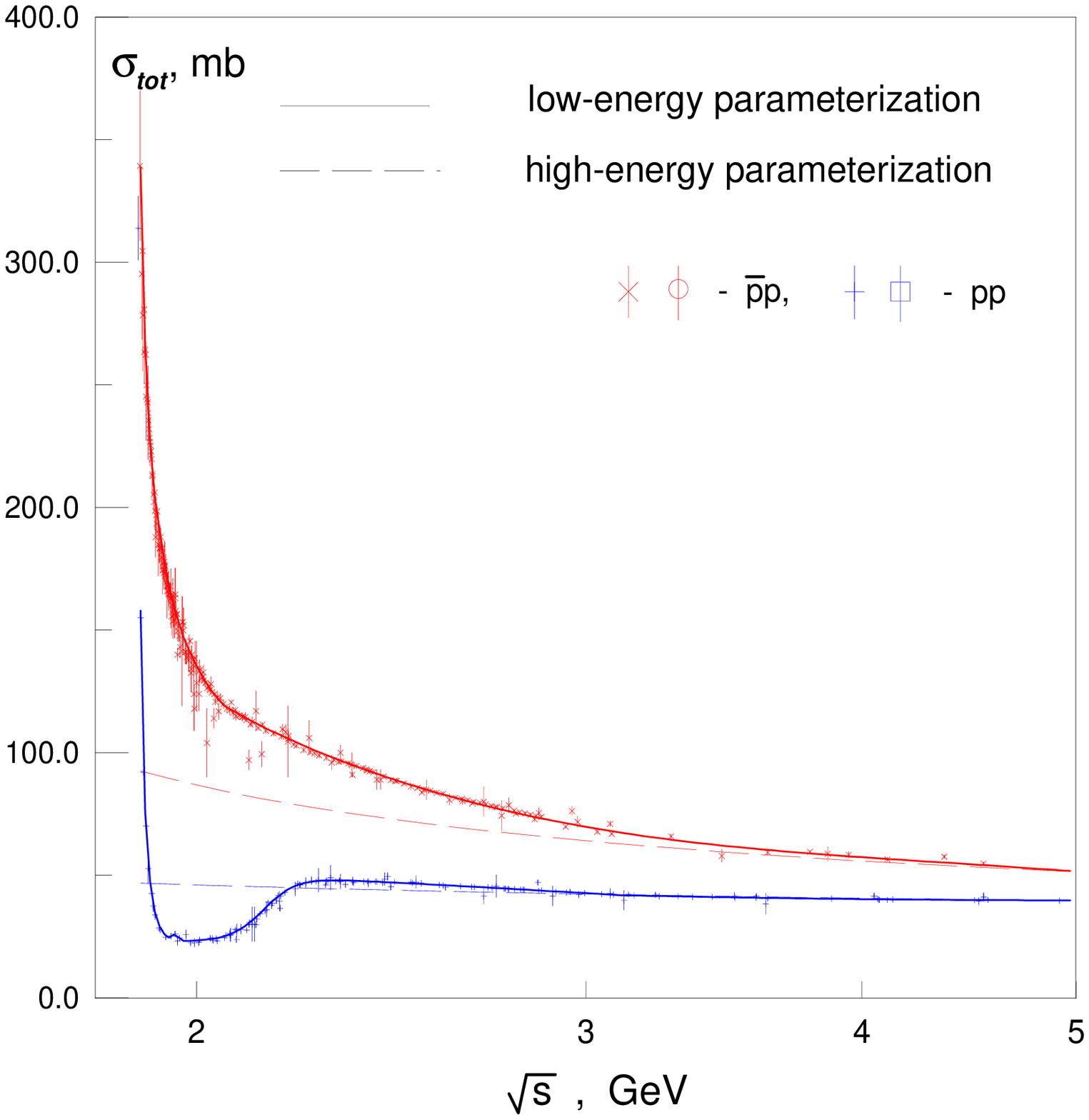}
\caption{Cross sections in the SP pomeron model at all energies (left) and
at at low energies (right). The solid lines are the cross-sections tuned
to SP model as described in the text. The dashed lines are the
continuation of the high-energy SP parameterization to lower energies.}
\label{fig:sigt}
\end{figure}

\begin{figure}[h]
\centering
\includegraphics[scale=0.45]{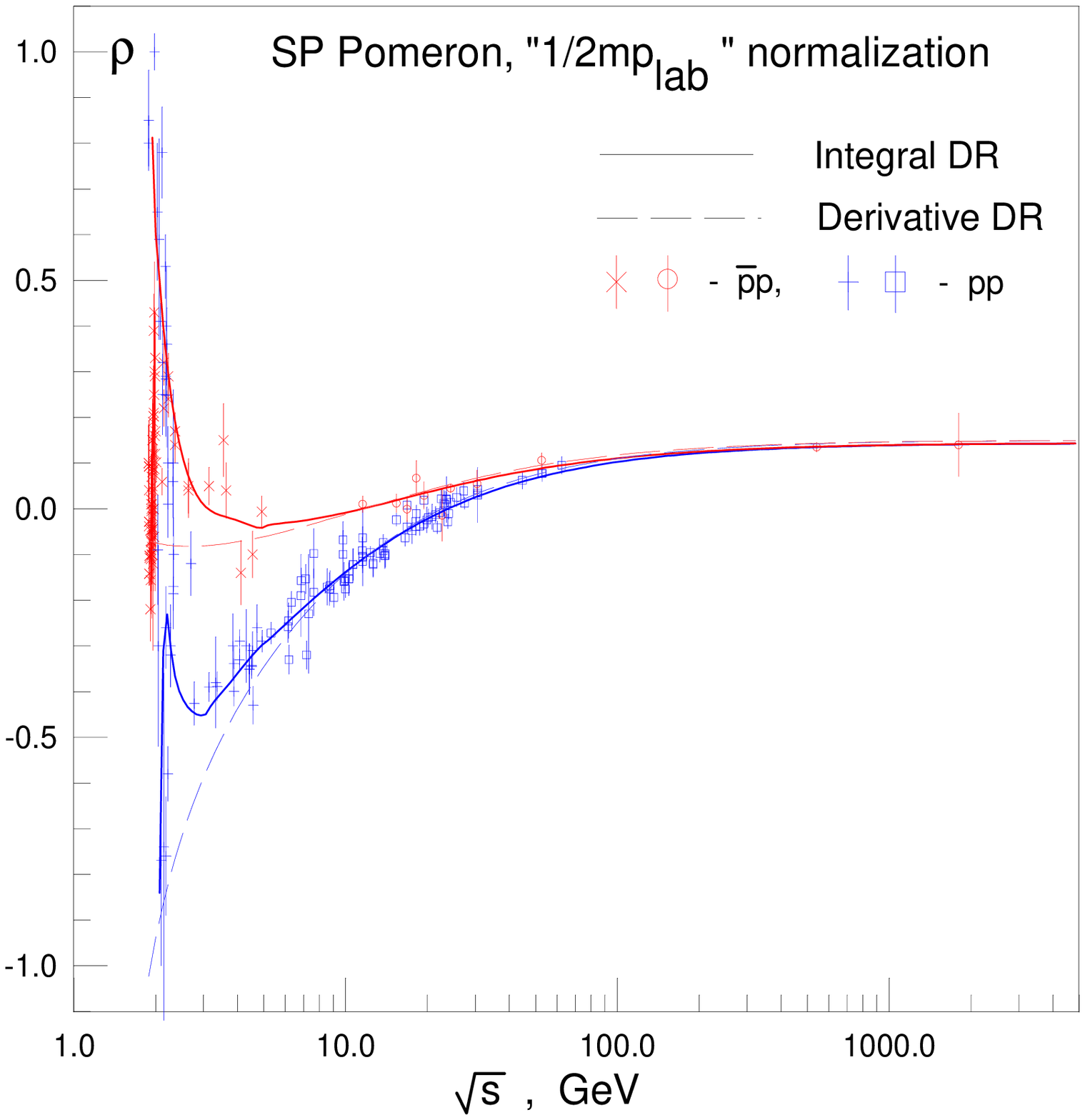}
\includegraphics[scale=0.45]{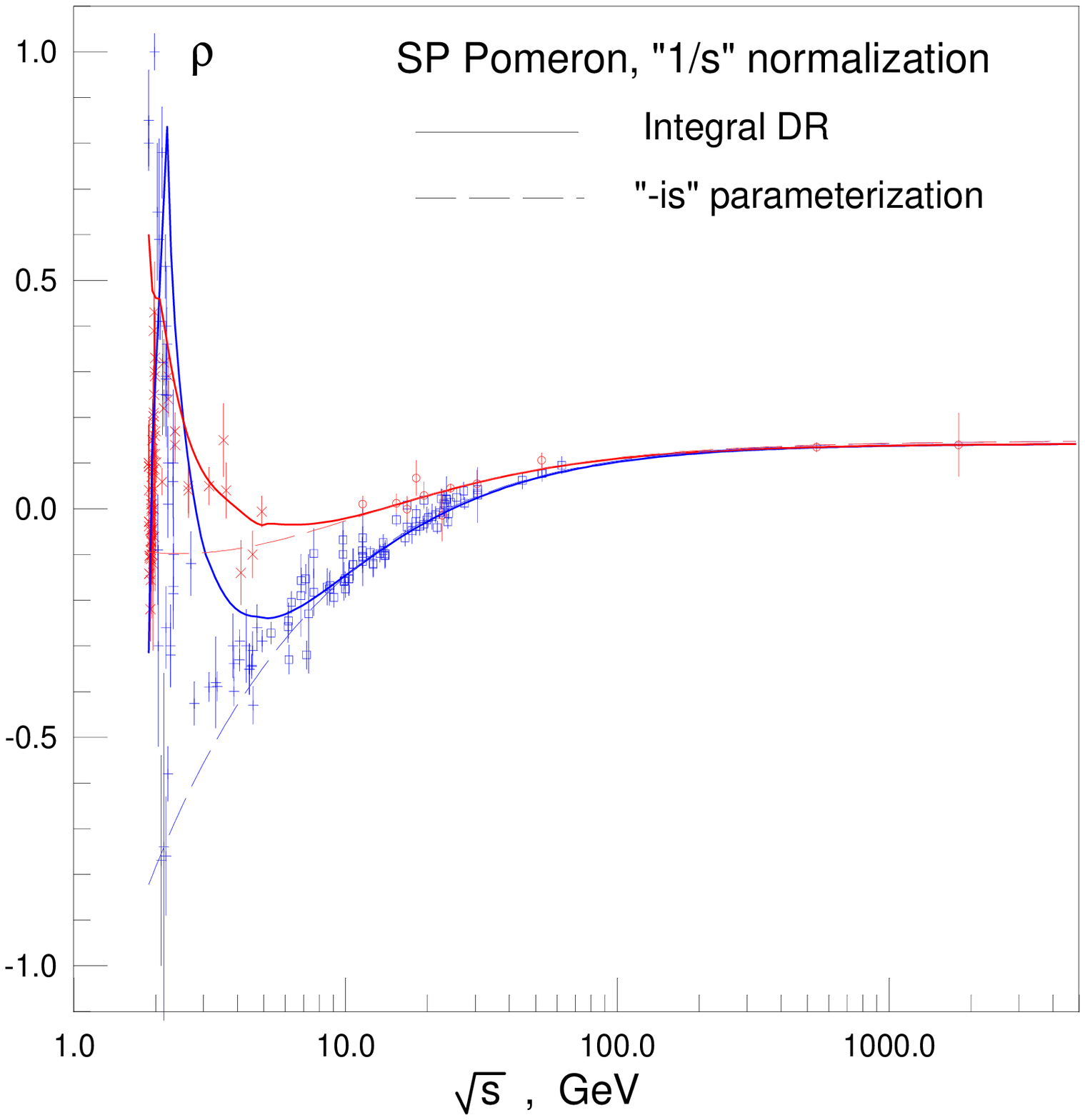}
\caption{The ratios $\rho$ in the SP model. Left: the integral and
derivative dispersion relation fit with the standard normalization. Right:
the IDR fit with asymptotic normalization and "$-is$"-fit (see text for
details).} \label{fig:d-lrho}
\end{figure}

We would like to insist on the fact that the values of $\rho$
calculated using DDR deviate from those calculated with IDR even at
$\sqrt{s} \lesssim 7 -8$ GeV (see Figs. 3, 4). It means that in order to
have more correct values of the $\rho$ at such energies, one must use the
IDR rather than explicit analytical expressions from the DDR.

The neglect of the subtraction constant, together with the use of
asymptotic formulae in a non-asymptotic domain, may be the source of the
conclusion of \cite{COMPETE} excluding the simple-pole model from the
list of the best models. Inclusion of these (non-asymptotic) terms
improves the description of $\rho$ considerably, and may lead to different
conclusions regarding the simple-pole model.

However, in order to have these final conclusions, one will have to make a
complete (a la COMPETE) analysis of the all data, including cross sections
and $\rho$ for $\pi p$ and $Kp$ interactions.


\begin{thebibliography}{}
\bibitem{rhohist}
M.L. Goldberger, Y. Nambu, R. Oehme, Ann. Phys. {\bf 2}, 226 (1957);\\
P. S\"{o}ding, Phys. Lett. {\bf 8}, 285, (1964);\\
M.M. Block, R.N. Cahn, Rev. Mod. Phys. {\bf 57},563 (1985);\\
U. Amaldi {\it et al.}, Phys. Lett. {\bf B66}, 390 (1985);\\
P. Kroll, W. Schwenger, Nucl. Phys. {\bf A503}, 865 (1989);\\
C. Aguer {\it et al.} (UA4 Collab.), Phys. Lett. {\bf B315}, 503,
(1993);\\
P. Desgrolard, M. Giffon, E.S. Martynov, Nuovo Cimento, {\bf 110A}, 537
(1997).
\bibitem{DDR1}
V.N. Gribov, A.A. Migdal, Yad. Fiz. {\bf 8}, 1002 (1968).
\bibitem{DDR2}
J.B. Bronzan,  G.L. Kane, U.P. Sukhatme, Phys. Lett. {\bf B49}, 272
(1974).
\bibitem{KN}
K. Kang, B. Nicolescu, Phys. Rev. {\bf D11}, 2561 (1975).
\bibitem{COMPETE}
J.R. Cudell {\it et al.} (COMPETE Collab.), Phys. Rev. {\bf D65}, 074024
(2002);\\
J.R. Cudell {\it et al.} (COMPETE Collab.), Phys. Rev. Lett. {\bf 89},
201801 (2002),
\bibitem{valeng}
A.I. Lengyel, V.I. Lengyel, Yad. Fiz. {\bf 11}, 669 (1970).
\bibitem{d-l}
A. Donnachie, P.V. Landshoff, Phys. Lett. {\bf B296}, 227 (1992).
\bibitem{data}
http://wwwppds.ihep.su:8001/hadron.html
\end{thebibliography}
\end{document}